\documentclass[a4,reqno,10pt]{amsart}

\usepackage[foot]{amsaddr}

 \usepackage{amsmath,amsthm,amssymb,xcolor,graphicx,soul}
 
\usepackage{float}

\usepackage{amscd,verbatim,a4wide}
\usepackage[all]{xy}

\usepackage[hidelinks]{hyperref}

\def\be{\begin{equation}}
\def\ee{\end{equation}}

\theoremstyle{plain}

\theoremstyle{definition}

\theoremstyle{remark}

\numberwithin{equation}{section}
\numberwithin{theorem}{section}
\numberwithin{figure}{section}
\numberwithin{table}{section}

\setstcolor{red}

\usepackage{tikz}
\usetikzlibrary{arrows}
\usetikzlibrary{positioning}
\usepackage{amsmath}
\usepackage{amsfonts}
\usepackage{hyperref}
\usepackage{amsthm}

\newcommand{\pr}{\partial}

\usepackage[english]{babel}
\usepackage{blindtext}

\begin{document}

\title[Spherical Photon Orbits around a 5D Myers-Perry Black Hole]{Spherical Photon Orbits around a 5D Myers-Perry Black Hole}

\author[M Bugden]{Mark Bugden}
\address[M Bugden]{Mathematical Sciences Institute,
Australian National University, 
Canberra, ACT 2601, Australia}
\email{mark.bugden@anu.edu.au}

\begin{abstract}
We study the motion of bound null geodesics with fixed coordinate radius around a five-dimensional rotating black hole. These spherical photon orbits are not confined to a plane, and can exhibit interesting quasiperiodic behaviour. We provide necessary conditions for the existence of these orbits, and explicitly compute the radii of circular orbits in the equatorial and polar planes. Finally, we plot representative examples of some of the types of possible orbits, commenting on their qualitative features.

\end{abstract}

\maketitle

 \section{Introduction}

 An interesting and well-known feature of the Schwarzschild black hole is the existence of a photon sphere - that is, a radius at which light can orbit the black hole. In this, and in more complicated black holes, the study of the motion of bound null geodesics around a black hole provides valuable information on the geometry of the spacetime. In four dimensions, the optical appearance of a star undergoing gravitational collapse \cite{AT68}, the shadow of the black hole \cite{MPO17}, how the night sky would look to an observer near a  black hole \cite{N93}, as well as quasinormal modes in the eikonal limit \cite{M85,CMBWZ09} are all related to the properties of unstable circular null geodesics. Spherical null geodesics are orbits which have a fixed cordinates radius and include, as a special case,  circular null geodesics. \\
 
 The study of spherical \emph{timelike} orbits around a Kerr black hole was initiated by Wilkins in \cite{W72}. Numerical integration in \cite{G74} produced an explicit example of such an orbit. Spherical \emph{null} geodesics around a Kerr black hole were studied by Teo in \cite{T03}. Teo's paper plotted several numerical examples of spherical photon orbits which were not confined to a plane. The interesting orbital dynamics of these geodesics demonstrates the complicated nature of photon spheres around rotating black holes. Recent work \cite{LDJ17} has studied similar orbits in an $\eta$-deformed Kerr black hole. \\
 
 Higher dimensional black holes arise naturally in the context of string theory and brane-world scenarios. These black holes are incredibly interesting objects to study from a theoretical physics perspective. The first string theoretic calculation of black hole entropy was done for a five-dimensional black hole \cite{SV96}, the gauge/gravity correspondence relates five-dimensional gravity to a four-dimensional QFT \cite{M99}, and higher dimensional black holes have been used to construct metrics on compact Sasaki-Einstein manifolds \cite{CHPP05}. Furthermore, higher dimensional black holes exhibit novel features not present in four dimensions, such as non-spherical horizon topologies \cite{ER02}, and violations of uniqueness results proven for four dimensional black holes \cite{EF07}. The Myers-Perry black holes are a very natural extension of the Kerr black hole to higher dimensional spacetimes \cite{MP86}. Although they have a spherical horizon topology, the behaviour of these black holes differs significantly from their four-dimensional counterparts. For example in five dimensions the Myers-Perry black holes can rotate in two independent planes of rotation, and in six and higher dimensions it is possible to have rotating black holes with arbitrarily high angular momentum in one of these planes without introducing a naked singularity. Such novel features provide a rich testing ground for mathematical relativity, and these solutions provide interesting and non-trivial objects to study in higher dimensional theories of gravity. \\
 
The aim of this paper is to extend the results of \cite{T03} to the class of five-dimensional Myers-Perry black holes, identifying the existence of bound null geodesics of constant coordinate radius and studying their qualitative features. The paper is organised as follows: In Section \ref{sec:EOM} we discuss the 5D Myers-Perry black hole solution, as well as the associated null geodesic equations. In Section \ref{sec:spherical} we discuss the conditions that need to be satisfied for spherical photon orbits. We also provide an explicit computation of the radii for the equatorial and polar null orbits around a generic 5D Myers-Perry black hole. In Section \ref{sec:examples} we present selected examples of spherical photon orbits obtained by numerical integration, and comment on their qualitative features. Finally, in Section \ref{sec:summary}, we summarise our results.  \\ 
 
 \section{Equations of Motion}\label{sec:EOM}
 The 5D Myers-Perry metric describes the geometry of a black hole in a five-dimensional spacetime, rotating in two independent planes. The line element of the 5D Myers Perry black hole in Boyer-Lindquist coordinates takes the following form:
 \begin{align}
 \label{MPlineelement}
 ds^2 &= -dt^2 + \frac{\mu}{\rho^2} \left[ dt + a \sin^2 \theta d\phi  + b \cos^2 \theta d\psi  \right]^2  +\frac{\rho^2}{4\Delta} dx^2 + \rho^2 d\theta^2 \\
 &\quad \quad + (x+a^2) \sin^2 \theta d \phi^2 + (x+b)^2 \cos^2 \theta d \psi^2 \notag,
 \end{align}
 where
 \begin{align*}
 \rho^2 &= x + a^2\cos^2\theta +b^2 \sin^2 \theta \\
 \Delta &= (x + a^2)(x + b^2) - \mu x.
 \end{align*}
 Here, the quantities $\mu$, $a$ and $b$ are quantities related to the mass $M$ and angular velocities $\Omega_a, \Omega_b$ in the following way:
 \begin{align*}
 M &= \frac{3\mu}{8\sqrt{\pi}G} \\
 \Omega_a &= \frac{a}{x_+ + a^2}\\
 \Omega_b &= \frac{b}{x_+ + b^2}.
 \end{align*}
 Note that following \cite{FS03}, the coordinate $x=r^2$ is used instead of the radius $r$ to simplify expressions.
 An event horizon of the black hole is located at $x_+$, where $$x_{\pm} = \frac{1}{2}\left(  \mu - a^2 - b^2 \pm \sqrt{(\mu - a^2 - b^2)^2 - 4a^2 b^2} \right).$$ This horizon exists and is positive provided $\mu \geq (a + b)^2$. In this paper, we shall only be interested in the region of the black hole exterior to this horizon, i.e. the region $x_+ < x < \infty$. \\
 The metric is invariant under the following symmetry transformation:
 \begin{align}
 \label{symmetry}
 a \longleftrightarrow b, \qquad \theta \longleftrightarrow \left( \frac{\pi}{2} - \theta \right), \qquad \phi \longleftrightarrow \psi.
 \end{align}
 The five first-order geodesic equations governing the motion of lightlike particles in this spacetime are derived in \cite{FS03}. They are
 \begin{subequations}
 \label{geodesic}
 \begin{align}
 \rho^2 \dot{t} &= E \rho^2  + \frac{\mu (x+a^2)(x+b^2)}{\Delta}\mathcal{E} \\
 \label{xeom}
 \left( \rho^2 \dot{x} \right)^2 &= 4 \mathcal{X} \\
 \label{geodesictheta}
 \left(\rho^2 \dot{\theta} \right)^2 &= \Theta \\
 \label{geodesicphi}
 \rho^2 \dot{\phi} &= \frac{\Phi}{\sin^2 \theta} - \frac{\mu a (x+ b^2)}{\Delta}\mathcal{E} - \frac{(a^2-b^2)\Phi}{x+a^2} \\
 \label{geodesicpsi}
 \rho^2 \dot{\psi} &=\frac{\Psi}{\cos^2 \theta} - \frac{\mu b (x+ a^2)}{\Delta}\mathcal{E} + \frac{(a^2-b^2)\Psi}{x+b^2},
 \end{align}
 \end{subequations}
 where the overdot represents differentiation with respect to an affine parameter along the geodesic, and the parameters $\{p_t = -E,\, p_{\phi} = \Phi,\, p_{\psi} = \Psi\}$ are the conserved quantities associated to the Killing vectors $\{\partial_t, \, \partial_{\phi}, \, \partial_{\psi}\}$. The functions $\mathcal{X}$ and $\Theta$ are given by:
 \begin{align*}
 \mathcal{X} &= \Delta \left[ x E^2 + (a^2 - b^2)\left( \frac{\Phi^2}{x+a^2} - \frac{\Psi^2}{x+b^2} \right) - K \right] + \mu (x+a^2)(x+b^2)\mathcal{E}^2 \\
 \Theta &= E^2(a^2 \cos^2 \theta + b^2 \sin^2 \theta) + K - \frac{\Phi^2}{\sin^2 \theta} - \frac{\Psi^2}{\cos^2 \theta}.
 \end{align*}
 The constant $K$ is related to Carter's constant, familiar from the study of geodesics around the four-dimensional Kerr black hole, and $\mathcal{E}$ is defined by 
 \begin{align*}
 \mathcal{E} &= E + \frac{a\Phi}{x+a^2} + \frac{b\Psi}{x+b^2}.
 \end{align*}

 For convenience, we use the invariance of geodesics under affine reparametrisation to `gauge away' the parameter $E$ by rescaling the affine parameter $\lambda \to E \lambda$. We then redefine the conserved quantities $\Phi, \, \Psi,$ and $K$ to be original quantities, divided by the appropriate power of $E$. In practice, this has the effect of setting $E = 1$.\\
 
 Note that the authors of \cite{DKLR14} find analytical solutions to these geodesic equations in terms of the Weierstrass $\wp$, $\zeta$, and $\sigma$ functions. Their analysis studies the general behaviour of geodesics by studying the zeros of $\Theta$ and $\chi$ (similar to our analysis in Section  \ref{subsec:eomtheta} and Section \ref{subsec:eomx}). From their analysis, they determine the existence of unstable circular bound orbits outside the event horizon. The orbits considered in this paper are a non-trivial generalisation of these circular orbits. \\

 \section{Conditions for spherical orbits}
 \label{sec:spherical}
 \subsection{The geodesic equation for $\theta$}
 \label{subsec:eomtheta}
 From (\ref{geodesictheta}), we see that the physically allowed range for $\theta$ occur when $\Theta \geq 0$, and we wish to find the values of the parameters for which this occurs. The relevant parameters are $a$ and $b$, describing the rotation of the black hole, and three constants of motion, $\Phi$, $\Psi$, and $K$.
 
 We first note that if $\Phi$ is non-zero, then $\Theta$ is negative and divergent as $\theta$ approaches $0$ or $\pi$. Similarly, if $\Psi$ is non-zero, then $\Theta$ is negative and divergent as $\theta$ approaches $\frac{\pi}{2}$. We conclude that orbits with non-zero $\Phi$ will be excluded from a region containing the poles, and orbits with non-zero $\Psi$ will be excluded from a region containing the equator. For an orbit with both $\Phi$ and $\Psi$ non-zero, $\theta$ is confined to a region $[\theta_- , \theta_+] \subset (0, \frac{\pi}{2})$ or $ (\frac{\pi}{2}, \pi)$. It follows that for any such orbit the expression
 \begin{align*}
 - \frac{ \Phi^2}{\sin^2 \theta} - \frac{ \Psi^2}{\cos^2 \theta}
 \end{align*} 
 is bounded below, and so the condition $\Theta \geq 0$ is always satisfied for large enough $K$. 
 
 Let us analyse the geodesic equation for $\theta$, with a view to constraining the physically allowed range of parameters. As we have already mentioned, for allowed orbits we require $\Theta \geq 0$. Since $\Theta$ is negative near $\theta =0$ and $\theta = \frac{\pi}{2}$, we require the existence of at least one root $\theta_0 \in (0,\frac{\pi}{2})$. We will consider the cases $a = b$ and $a \not= b$ separately. 
 \subsubsection{$a = b$}
 When the rotation parameters are equal, the geodesic equation for $\theta$ simplifies to 
 \begin{align}
 \Theta = a^2 + K - \frac{\Phi^2}{\sin^2 \theta} - \frac{\Psi^2}{\cos^2 \theta}.
 \end{align}
 This function has a maximum at $\theta = \tan^{-1} \left( \sqrt{\frac{\Phi}{\Psi}} \right)$, where it attains the value $\Theta_{max} = a^2 + K - (\Phi + \Psi)^2$. It follows that $\Theta$ has at least one root precisely when $K \geq (\Phi+\Psi)^2 -a^2$.
 \subsubsection{$a \not= b$}
 When the rotation parameters are not equal, the analysis is a bit more complicated. We are able to find necessary conditions for the existence of at least one root, although they are not sufficient. We begin by assuming that $a^2 < b^2$. We also assume that the angular momenta are generic, that is, $\Phi \not = 0$ and $\Psi\not = 0$. We will discuss these special cases later. If we set $y = \sin^2 \theta$, then we can rewrite $\Theta$ as 
 \begin{align*}
 \Theta(y) &= \frac{1}{y(1-y)} \left[(a^2-b^2) y^3 + (b^2-2a^2 - K) y^2  + (\Phi^2 - \Psi^2 + a^2 + K) y - \Phi^2\right] \\
 &= \frac{Q(y)}{y(1-y)}.
 \end{align*}
 With this simplification, we are now looking for roots of the polynomial $Q(y)$ with $y \in (0,1)$. To study roots in this interval, we will use the Fourier-Budan theorem \cite{B1807,F1820}, the pertinent details of which we now recall. For a polynomial $p(x)$ with real coefficients, let $\mathcal{F}_p(x) = \{ p(x), p'(x), p''(x),\dots\}$ be the corresponding Fourier sequence, and let $v_p(x)$ be the number of sign changes\footnote{if a term in the sequence is zero, it is omitted} in the sequence $\mathcal{F}_p(x)$. The Fourier-Budan theorem then asserts two results:
 \begin{itemize}
 	\item $v_p(0) \geq v_p(1)$
 	\item The number $\rho$ of real roots of the polynomial $p(x)$ located in the open interval $(0,1)$ satisfies $\rho \leq v_p(0) - v_p(1)$.
 \end{itemize} 
 In particular, if there are no sign changes in $\mathcal{F}_p(0)$, then there are no real roots of $p(x)$ in the interval $(0,1)$.
 The relevant Fourier sequences for $Q(y)$ are
 \begin{align*}
 \mathcal{F}_Q(0) &= \{ -\Phi^2,\, \Phi^2 - \Psi^2 + a^2 + K ,\, 2b^2 - 4a^2 -2K,\, 6a^2 - 6b^2 \}\\
 \mathcal{F}_Q(1) &= \{ -\Psi^2,\, \Phi^2 - \Psi^2 - b^2 - K ,\, 2a^2 - 4b^2-2K,\, 6a^2-6b^2\}.
 \end{align*}
 Since $a^2 < b^2$, we see immediately that the first and last terms of $\mathcal{F}_Q(0)$ are negative. If all the terms of $\mathcal{F}_Q(0)$ are negative, then $v_Q(0) = 0$, and it follows that there are no real roots in the interval $(0,1)$. Thus, a necessary condition for the existence of positive roots in $(0,1)$ is that at least one of the middle terms in $\mathcal{F}_Q(0)$ should be positive. This gives the following constraints on $K$:
 \begin{align*}
 K < b^2 - 2a^2 \qquad \textrm{or} \qquad  K > \Psi^2 - \Phi^2 - a^2.
 \end{align*}
 Note that if $a^2 >b^2$,we can do a similar analysis by replacing the substitution $y = \sin^2 \theta$ with $y = \cos^2 \theta$. Alternatively, we can simply utilise the symmetry (\ref{symmetry}) of the metric. Either way, the constraints we obtain are:
 \begin{align}
  K < a^2 - 2b^2 \qquad \textrm{or} \qquad  K > \Phi^2 - \Psi^2 - b^2.
 \end{align}
 \subsection{The geodesic equation for $x$}
 \label{subsec:eomx}
 We are interested in spherical photon orbits with constant radius, $x$, so we want to solve the conditions $\mathcal{X} = \frac{d\mathcal{X}}{dx} = 0$ at this radius. That is, we are looking for a zero of order 2 for the equation $\chi(x)$ with $x >x_+$. 
 Although $\chi$ looks intimidating, it reduces to a relatively simple cubic equation
 \begin{align*}
 \chi = a_3 x^3 + a_2 x^2 + a_1 x + a_0
 \end{align*}
 with
 \begin{align*}
 a_3 &= 1 \\
 a_2 &= a^2+b^2-K \\
 a_1 &= (\Phi^2-\Psi^2)(a^2-b^2) + K(\mu - a^2 - b^2)\\
 & \quad  + a^2 b^2 + \mu (a^2 + b^2 + 2a\Phi + 2b \Psi)\\
 a_0 &= \Phi^2 b^2 (a^2 - b^2 + \mu) + \Psi^2 a^2 (b^2-a^2 + \mu) \\ 
 & \quad - K a^2 b^2 + a b \mu (ab + 2 \Phi b + 2 \Psi a + 2 \Phi \Psi).
 \end{align*}
 Following the example of \cite{T03}, we can solve the equations $\chi = 0$ and $\frac{d \chi}{d x} = 0$ simultaneously for two of the given parameters, such as $x$ and $K$. The families of solutions obtained are very complicated, so we don't reproduce them here, although they can be computed with a computer software package such as Mathematica or Maple. In practice, we first decide on a particular choice of black hole, corresponding to a choice of the parameters $a,\, b,\, \mu$, and then select appropriate values of $\Phi$ and $\Psi$. Substituting these values into the solutions for $x$ and $K$ then gives us the full set of parameters satisfying $\chi = \frac{d \chi}{dx} = 0$. Recall that a cubic equation has a double root if and only if the discriminant $$D = 18a_3 a_2 a_1 a_0 -4a_2^3 a_0 + a_2^2 a_1^2 - 4 a_3 a_1^3 - 27 a_3^2 a_0^2$$ vanishes. Once a particular set of parameter values are chosen, the discriminant gives us a quick way to verify that they do indeed solve the equations $\chi = \frac{d \chi}{d x} = 0$. Of course, these parameters must also satisfy the constraints imposed by the geodesic equation for $\Theta$. \\
 
In \cite{FS03}, it was shown that there are no stable circular orbits in the equatorial plane. The authors of \cite{DKLR14} argue that for certain values of the parameters, there are no stable bound orbits outside the event horizon. Whilst we have been unable to prove this for arbitrary values of the parameters, it should be noted that all of the explicit examples of orbits we discuss have $\frac{d^2 \chi}{ d x^2} >0$, and are therefore unstable under radial perturbations. 

\subsection{Rewriting the geodesic equations}
The null geodesic equations, (\ref{geodesic}), have the unfortunate feature that the geodesic equation for $\theta$ is quadratic in $\dot{\theta}$. This is an inconvenience during numerical integration since one must manually change the sign at turning points of the function. To circumvent this issue, we follow the procedure of \cite{LDJ17} by converting to the Hamiltonian formulation, where the resulting equations are not quadratic. To begin, we will need the following expressions of the components of the inverse metric:
\begin{subequations}
\label{inversemetric}
\begin{align}
\label{inversex}
g^{xx} &= \frac{4 \Delta}{\rho^2} \\
\label{inversetheta}
g^{\theta \theta} &= \frac{1}{\rho^2}\\
g^{tt} &= \frac{1}{\rho^2} \left[ (a^2-b^2)\sin^2 \theta - \frac{(x+a^2)[\Delta + \mu (x+b^2)]}{\Delta} \right]\\
g^{t \phi} &= \frac{a \mu (x+b^2)}{\rho^2 \Delta} \\
g^{t \psi} &= \frac{b \mu (x+a^2)}{\rho^2 \Delta} \\
g^{\phi \phi} &= \frac{1}{\rho^2} \left[ \frac{1}{\sin^2 \theta} - \frac{(a^2-b^2)(x+b^2)+b^2 \mu}{\Delta} \right] \\
g^{\psi \psi} &= \frac{1}{\rho^2}\left[ \frac{1}{\cos^2 \theta} + \frac{(a^2-b^2)(x+a^2)-a^2 \mu}{\Delta} \right]  \\
g^{\phi \psi} &= - \frac{ab \mu}{\rho^2 \Delta}.
\end{align}
\end{subequations}
The Hamiltonian is, as usual, defined by $H = \frac{1}{2} g_{\mu \nu} \dot{q}^{\mu} \dot{q}^{\nu} =  \frac{1}{2} g^{\mu \nu} p_{\mu} p_{\nu}$. Recalling that the momenta $\{p_t, p_{\phi}, p_{\psi}\}$ are constant along null geodesics, we have
\begin{align}
H = \frac{1}{2}g^{xx} p_x^2 + \frac{1}{2}g^{\theta \theta}p_{\theta}^2 + f(x,\theta,E,\Phi,\Psi),
\end{align}
for some as yet undetermined function $f$. We now use the relation $p_{\mu} = g_{\mu \nu} \dot{q}^{\nu}$ together with (\ref{geodesic}) to obtain
\begin{subequations}
\label{momenta}
\begin{align}
p_x^2 &= \frac{\chi}{4 \Delta^2} \\
p_{\theta}^2 &= \Theta.
\end{align}
\end{subequations}
The relations (\ref{momenta}) together with the null condition $g_{\mu \nu} \dot{q}^{\mu} \dot{q}^{\nu}=0$ then give us
\begin{align}
H = \frac{\chi}{2\rho^2 \Delta} + \frac{\Theta}{2 \rho^2} + f(x,\theta,E,\Phi,\Psi) = 0.
\end{align}
This fixes $f$ and lets us write the Hamiltonian as 
\begin{align}
H (q,p) = \frac{2 \Delta}{\rho^2}p_x^2 + \frac{1}{2 \rho^2} p_{\theta}^2 - \left( \frac{\chi + \Delta \Theta}{2 \rho^2 \Delta} \right).
\end{align}
We can now use Hamilton's equations to obtain
\begin{subequations}
\begin{align}
\dot{x} &= \frac{4 \Delta}{\rho^2} p_x \\
\dot{\theta} &= \frac{1}{\rho^2} p_{\theta} \\
\dot{t} &= \frac{1}{2 \rho^2 \Delta}\frac{\pr \left(\chi + \Delta \Theta\right)}{\pr E} \\
\dot{\phi} &= - \frac{1}{2 \rho^2 \Delta}\frac{\pr \left(\chi + \Delta \Theta\right)}{\pr \Phi} \\
\dot{\psi} &= - \frac{1}{2 \rho^2 \Delta}\frac{\pr \left(\chi + \Delta \Theta\right)}{\pr \Psi} \\
\dot{p}_x &= -\left( \frac{2 \Delta}{\rho^2}\right)_{\!\!,x} p_x^2 - \left( \frac{1}{2 \rho^2}\right)_{\!\!,x} p_{\theta}^2 + \left( \frac{\chi + \Delta \Theta}{2 \rho^2 \Delta}\right)_{\!\!,x} \\
\dot{p}_{\theta} &= -\left( \frac{2 \Delta}{\rho^2}\right)_{\!\!,\theta} p_x^2 - \left( \frac{1}{2 \rho^2}\right)_{\!\!,\theta} p_{\theta}^2 + \left( \frac{\chi + \Delta \Theta}{2 \rho^2 \Delta}\right)_{\!\!,\theta} 
\end{align}	
\end{subequations}
where the comma denotes differentiation with respect to that variable. These equations certainly look more complicated than (\ref{geodesic}), but they are more convenient for numerical integration.


\subsection{A note on equatorial orbits}
\label{subsec:equatorial}
Kerr black holes in four dimensions have two circular null geodesics in the equatorial plane, and Myers-Perry black holes also allow circular null orbits in the equatorial plane. For a singly spinning Myers-Perry black hole, the radii of such orbits was computed explicitly in \cite{CMBWZ09}. Here, we give a computation for the radii of circular null orbits in the equatorial plane of an arbitrary Myers-Perry black hole in five dimensions.\footnote{i.e. for generic $a$ and $b$}\\

To study circular orbits in the equatorial plane, we fix $\theta = \frac{\pi}{2}$. For simplicity, we also set $\mu = 1$. In order for such orbits to remain at constant $\theta$, we require that $\Theta = 0$. We see immediately from (\ref{geodesictheta}) that this is only possible if $\Psi = 0$. Solving $\Theta = 0$ for the parameter $K$ then gives us $K = \Phi^2 - b^2$. Substituting this value of $K$ into $\chi$ and $\frac{d \chi}{dx}$ give us the following cubic and a quadratic in $x$:
\begin{align*}
\chi &= x^3 + (a^2 + 2b^2 - \Phi^2) \, x^2 + (a^2 + \Phi^2 + 2 a \Phi + b^4 + 2 a^2 b^2 - 2 \Phi^2 b^2)\,x \\
& \quad + (a^2 + \Phi^2 + 2a \Phi + b^4 + 2 a^2 b^2 - 2 \Phi^2 b^2)\\
\frac{d \chi}{dx} &= 3x^2 + 2(a^2 + 2b^2 - \Phi^2) \, x + (a^2 + \Phi^2 + 2 a \Phi + b^4 + 2 a^2 b^2 - 2 \Phi^2 b^2).
\end{align*}
We then solve the equations $\chi = 0$ and $\frac{d \chi}{dx} = 0$ simultaneously for $x$ and $\Phi$. There are five solutions, one of which has $x = -b^2 < 0$, which we shall ignore. The remaining solutions are:
\begin{align}
x &= 1+a-b^2 \pm \sqrt{(1+a)^2 - b^2} \\
\Phi &= 1 \pm \sqrt{(1+a)^2-b^2}
\end{align}
and 
\begin{align}
x &= 1-a-b^2 \pm \sqrt{(1-a)^2 - b^2} \\
\Phi &= -1 \mp \sqrt{(1-a)^2-b^2}.
\end{align}
Note that the two values of $x$ which have positive signs in front of the square root are always greater than or equal to $x_+$ for all $(a+ b)^2<1$. Similarly, the values of $x$ which have negative signs in front of the square root are always less than or equal to $x_+$. Since we are looking for photon orbits of constant radius \emph{outside} the event horizons, we conclude there are exactly two such orbits for each value of $(a,\, b)$, with $0<(a+b)^2<1$. The radii of these orbits are
\begin{align}
x = 1 \pm a - b^2 + \sqrt{(1 \pm a)^2 - b^2}.
\end{align}
When $b = 0$, we obtain
\begin{align}
x = 2(1 \pm a),
\end{align}
in agreeance with the results of \cite{CMBWZ09}.\\
We can do a similar analysis to determine the radii of circular null orbits for constant $\theta = 0$. We spare the reader the details of the calculation, and simply include the result:
\begin{align}
x = 1 \pm b - a^2 + \sqrt{(1\pm b)^2-a^2}.
\end{align}

\section{Examples of selected orbits}
\label{sec:examples}

\subsection{Visualising the orbits}
A significant obstacle to studying the orbits of particles around five-dimensional black holes is the inability to visualise these orbits. We present here a few methods by which one can attempt to visualise such orbits. 

We are interested in null geodesics that have the property that their coordinate radius $r$ is constant. The orbit of such a geodesic can be described as a path in $\{\theta,\, \phi,\, \psi\}$ space, and this is the simplest method of visualising these orbits when you are able to view the three-dimensional plot. We can also plot the functions $\theta,\, \phi,\, \psi$ on the same graph, as functions of the affine parameter $\lambda$. This method is particularly useful when one is confined to two-dimensional plots. 

\begin{figure}[h!]
\includegraphics[width = 8.4cm]{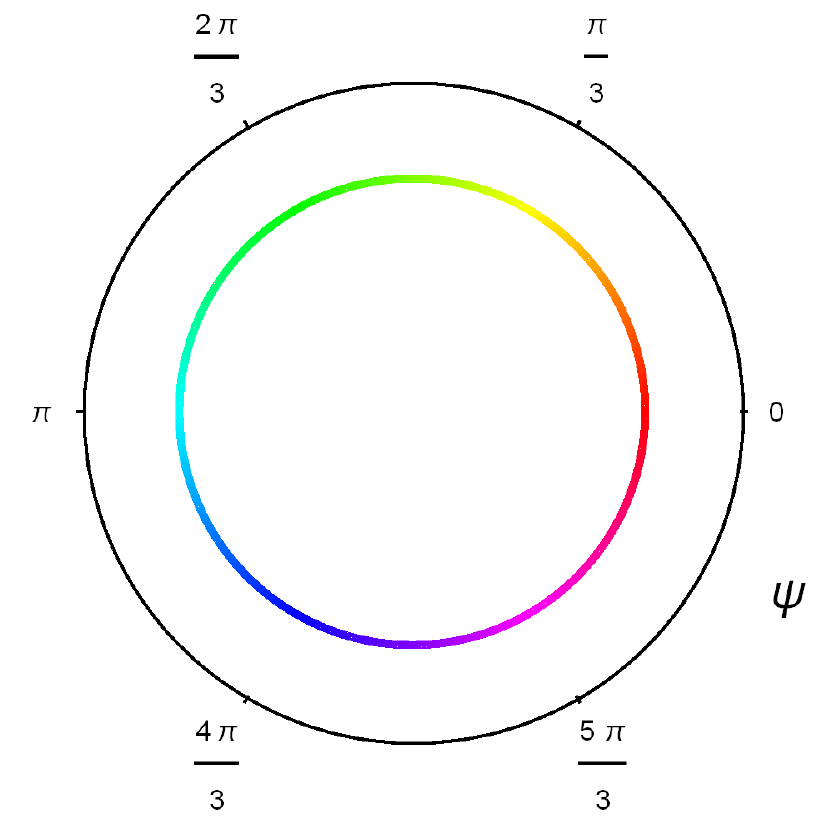}
\caption{The colour wheel mapping the coordinate $\psi$ to a colour}
\label{colorwheel}
\end{figure}

Another method involves a coordinate change to pseudo-cartesian coordinates. Following \cite{MP86}, we make the following change of coordinates
\begin{align*}
x_1 &= \sqrt{r^2 + a^2} \, \sin \theta \cos \left[ \phi - \arctan \left(\frac{a}{r} \right)\right] \\
y_1 &= \sqrt{r^2 + a^2} \, \sin \theta \sin \left[ \phi - \arctan \left(\frac{a}{r} \right)\right] \\
x_2 &= \sqrt{r^2 + b^2} \, \cos \theta \cos \left[ \psi - \arctan \left(\frac{b}{r} \right)\right] \\
y_2 &= \sqrt{r^2 + b^2} \, \cos \theta \sin \left[ \psi - \arctan \left(\frac{b}{r} \right)\right].
\end{align*} 
In these coordinates, the metric (\ref{MPlineelement}) takes the Kerr-Schild form
\begin{align}
g_{\mu \nu} &= \eta_{\mu \nu} + hk_{\mu} k_{\nu},
\end{align}
where $k_{\mu}$ is a null vector field and $h$ is a function of $\{r,\, x_i,\, y_i\}$. The substitution $z_1 = \sqrt{(x_2)^2 + (y_2)^2}$ then gives us the following three-dimensional coordinate system:
\begin{align*}
x_1 &= \sqrt{r^2 + a^2} \, \sin \theta \cos \left[ \phi - \arctan \left(\frac{a}{r} \right)\right] \\
y_1 &= \sqrt{r^2 + a^2} \, \sin \theta \sin \left[ \phi - \arctan \left(\frac{a}{r} \right)\right] \\
z_1 &= \sqrt{r^2+b^2} \, \cos \theta.
\end{align*}
This is reminiscent of the cartesian description of the four-dimensional Kerr black hole, albeit deformed by $b$. Surfaces of constant $r$, to which our orbits are confined, are spheroids.\footnote{These spheroids are \emph{oblate} if $a^2>b^2$ and \emph{prolate} if $a^2<b^2$. If $a^2 = b^2$, the surface is a sphere.} This projection to three dimensions is not without its drawbacks - we have lost the information contained in the $\psi$ coordinate. Since $\psi$ is an angular coordinate, we can map the coordinate to a colour wheel (see Figure \ref{colorwheel}). The orbit in the three-dimensional cartesian space $\{x_1,y_1,z_1\}$ can then be coloured according to the value of $\psi$, allowing us to visualise the orbit. 

\subsection{Explicit orbits}
In this section we present explicit examples of spherical photon orbits.

\subsubsection{One plane of rotation}

 We begin by considering black holes with one vanishing rotation parameter, that is with $b=0$. We are able to find qualitatively similar orbits to the spherical photon orbits found in \cite{T03}, as well as orbits which are qualitatively different. Note that in order to have an event horizon for this black hole, we require $a^2 < \mu $. We set $\mu = 1$ by fiat. \\

\begin{figure}[H]
	\includegraphics[width = 8.4cm]{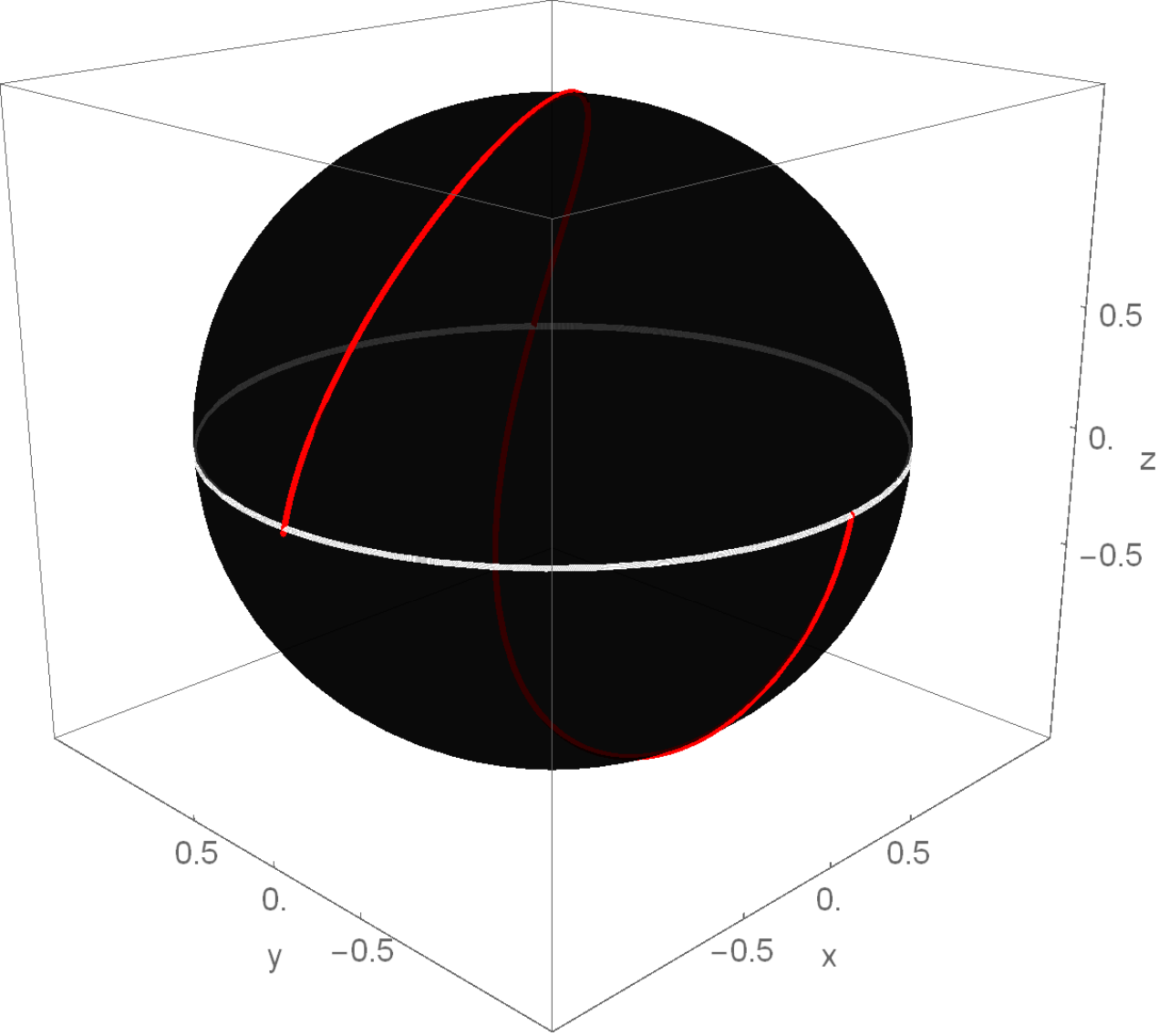}
	\caption{One latitudinal oscillation of the zero momentum orbit. The rotation parameter of the black hole is $a = 0.5$. }
	\label{fig:nomomenta}
\end{figure}

The first example (Figure \ref{fig:nomomenta}) is an orbit which moves through all possible latitudes. The angular momenta are both zero, and the radius of the orbit is fixed to $x = 2 \mu - a^2 \equiv x_+ + \mu$ by the constraints $\chi = \frac{d \chi}{dx} = 0$. We take $a = 0.5$ for simplicity, with other values being qualitatively similar. Like the case of the non-equatorial, zero-angular-momentum orbit around four-dimensional Kerr, this orbit is non-planar. Since $\dot{\psi} = 0$ by (\ref{geodesicpsi}), the angular coordinate $\psi$ will be constant along this geodesic. This orbit moves through all latitudes, periodically returning to the north and south pole. \\

\begin{figure}[h]
	\includegraphics[width = 8.4cm]{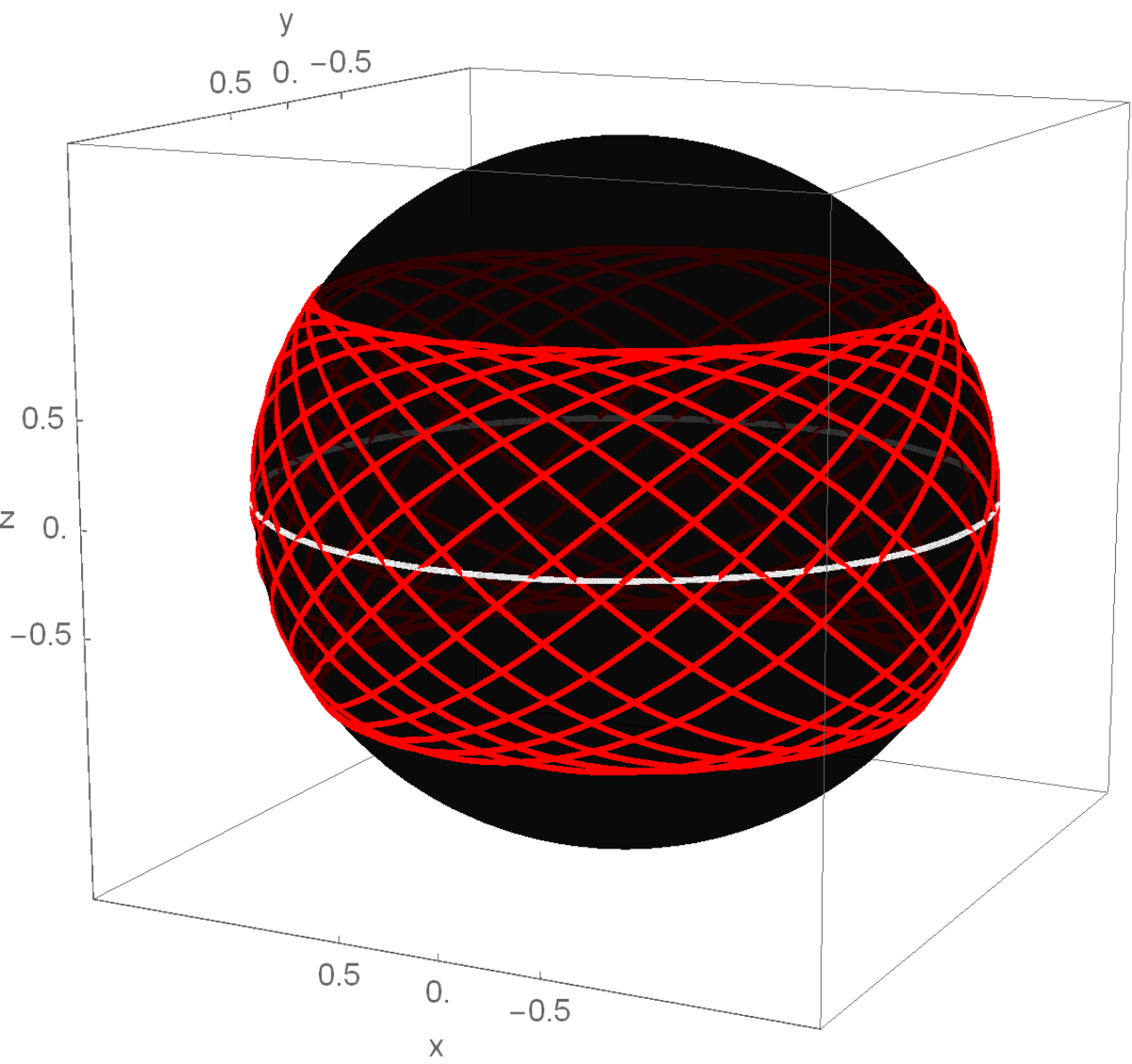}
	\caption{Many latitudinal oscillations of an orbit with $\Phi=2$. The rotation parameter of the black hole is $a = 0.5$. }
	\label{fig:phiequals2}
\end{figure}

The next example (Figure \ref{fig:phiequals2}) is another orbit of one of the types found in \cite{T03}, which has a maximum and a minimum latitude. As before, since both $b$ and $\Psi$ are vanishing, there is no motion in the $\psi$ direction. This example exhibits quasiperiodic behaviour, confined to a band around the equator. \\

\begin{figure}[h]
	\includegraphics[width = 8.4cm]{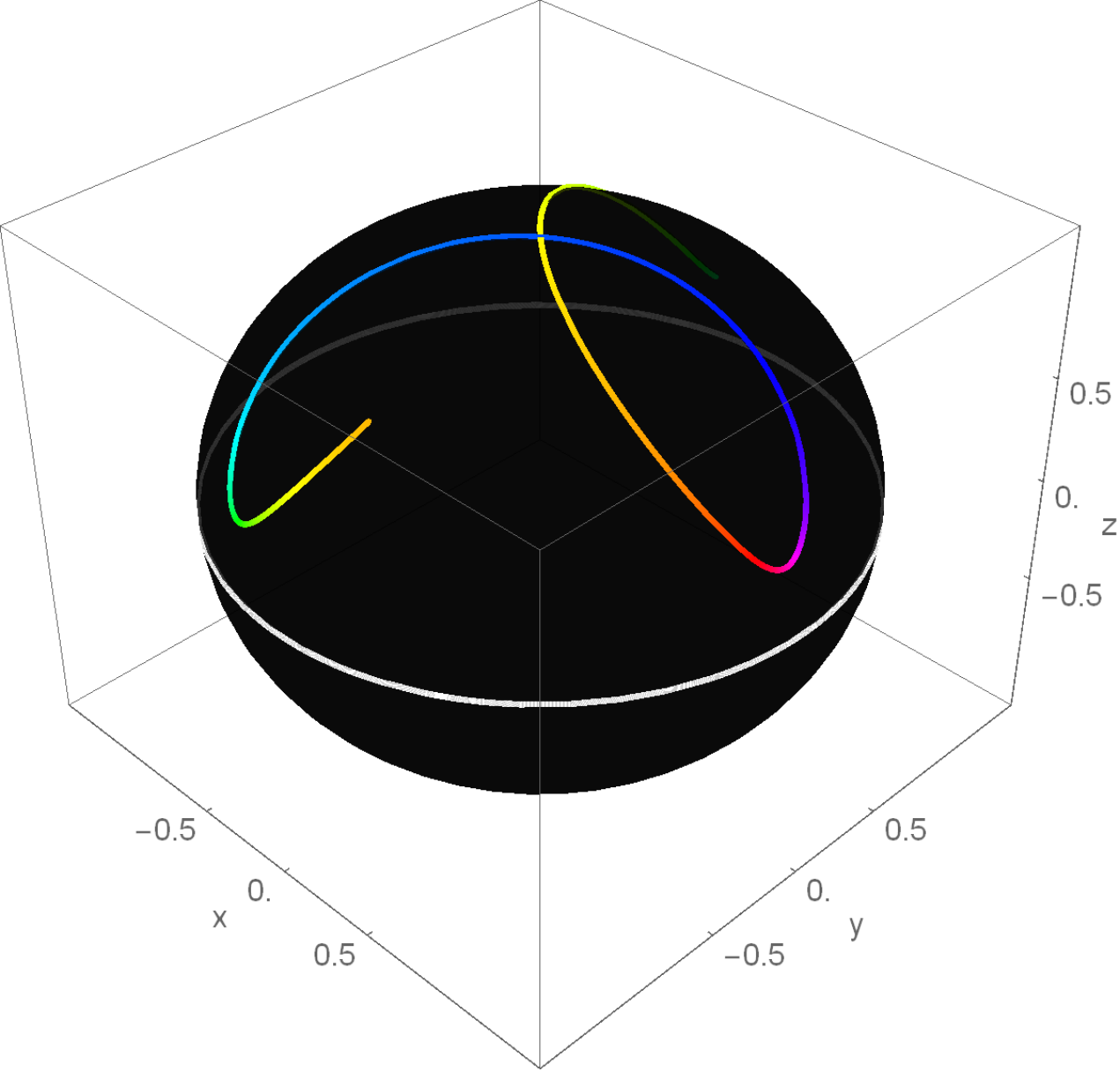}
	\caption{An orbit with $\Psi = 0.5$ and $\Phi = 0$. The rotation parameter of the black hole is $a = 0.8$. }
	\label{fig:firstcolour}
\end{figure}

The first two orbits are qualitatively similar to the four-dimensional Kerr orbits. The next example (Figure \ref{fig:firstcolour}), though, is a new type of orbit. This orbit has $\Phi = 0$, but it has a non-zero $\Psi$.

\begin{figure}[h]
	\includegraphics[width = 8.4cm]{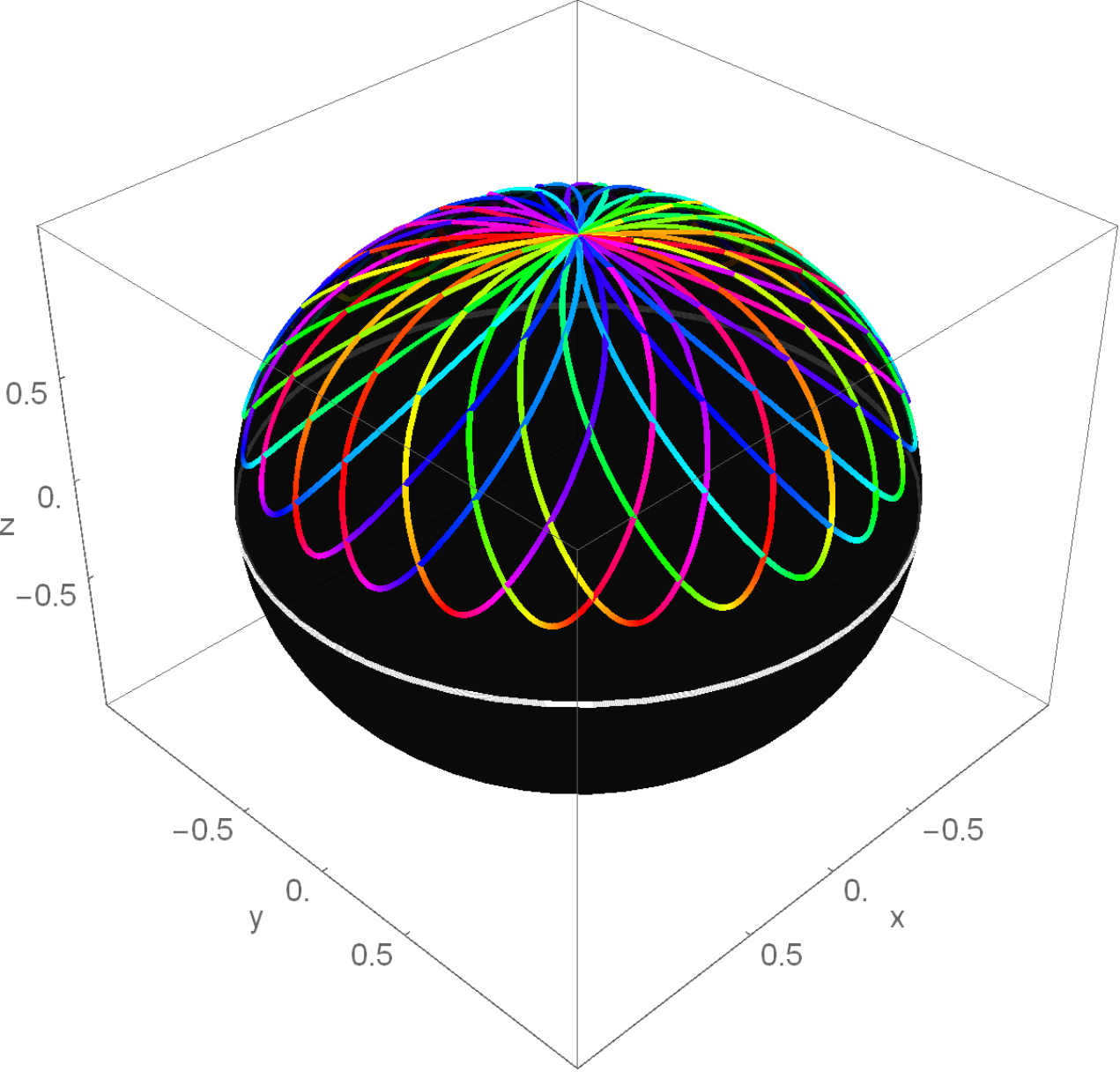}
	\caption{Many oscillations of an orbit with $\Psi = 0.5$ and $\Phi = 0$. The rotation parameter of the black hole is $a = 0.8$. }
	\label{fig:manycolour}
\end{figure}

Like the $\Phi=0$ orbit we saw earlier, this orbit periodically returns to the north pole at $\theta = 0$. This orbit, however, now has a \emph{minimum} latitude it can achieve, readily apparent from Figure \ref{fig:manycolour}. This is in agreeance with the results of Section \ref{subsec:eomtheta}, which states that orbits with a non-zero $\Psi$ are confined to regions which exclude the equator.\\

\begin{figure}[h]
	\includegraphics[width = 8.4cm]{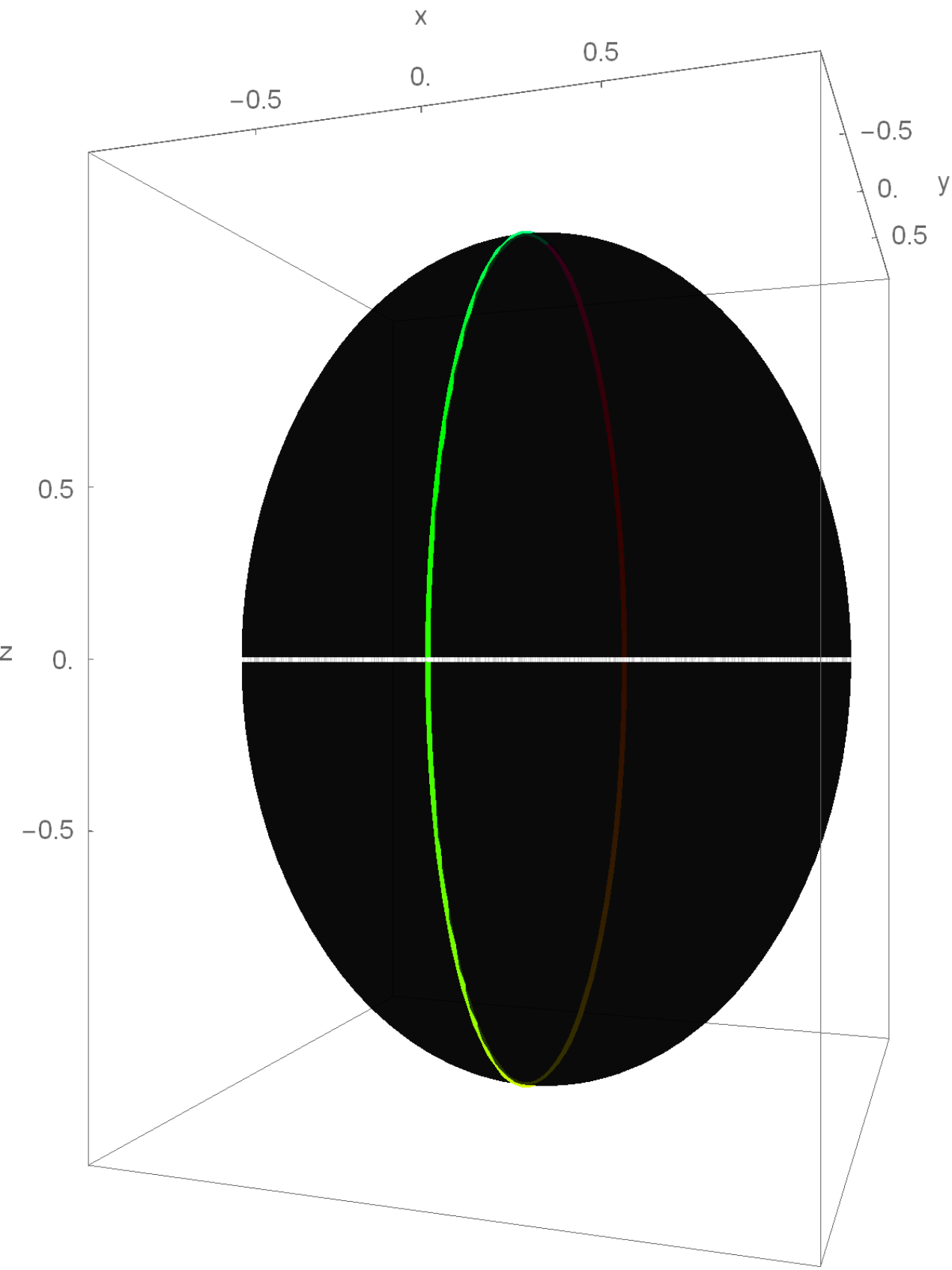}
	\caption{One latitudinal oscillation of the zero momenta orbit around the `football' black hole, with $a =0$ and $b = 0.99$. }
	\label{fig:Football}
\end{figure}

For the final example of this section, we look at an orbit around a black hole with vanishing $a$ and non-vanishing $b$. This black hole also allows a zero-momenta orbit, shown in Figure \ref{fig:Football}. Since $a$ and $\Phi$ vanish for this orbit, $\phi$ is constant by Equation (\ref{geodesicphi}). \\

Although this orbit looks qualitatively different to the orbit of Figure \ref{fig:nomomenta}, it is worth mentioning that they are actually the same type of orbit, related to each other by the symmetry transformation (\ref{symmetry}). The apparent difference arises from our method of visualising the orbits, which privileges the plane of rotation defined by $a$. 

\subsubsection{Two planes of rotation}

\begin{figure}[h]
	\includegraphics[width = 8.4cm]{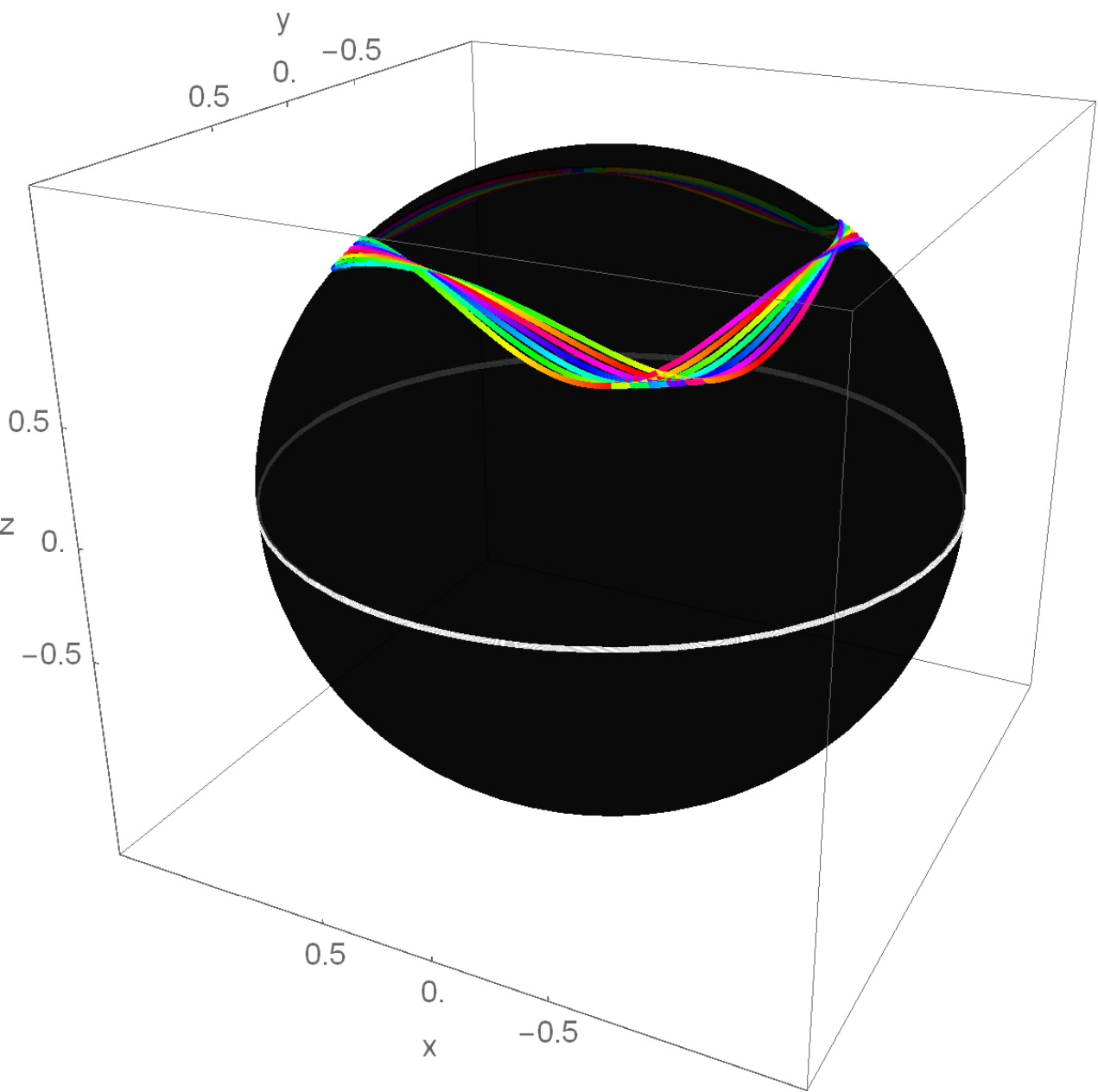}
	\caption{Many latitudinal oscillations of an orbit with $\Phi = 1$, $\Psi = -1$, around a black hole with rotation parameters $a = 0.6$ and $b = 0.3$.}
	\label{fig:Band}
\end{figure}

In this section, we will exhibit examples of orbits around Myers-Perry black holes with two non-vanishing rotation parameters. The first example (Figure \ref{fig:Band}) in this section has a non-zero $\Phi$ and $\Psi$, so by the analysis in Section \ref{subsec:eomtheta}, this orbit is confined to a region $[\theta_- , \theta_+] \subset (0, \frac{\pi}{2})$, combining the features of Figure \ref{fig:phiequals2} and Figure \ref{fig:manycolour}.\\

\begin{figure}[h]
	\includegraphics[width = 8.4cm]{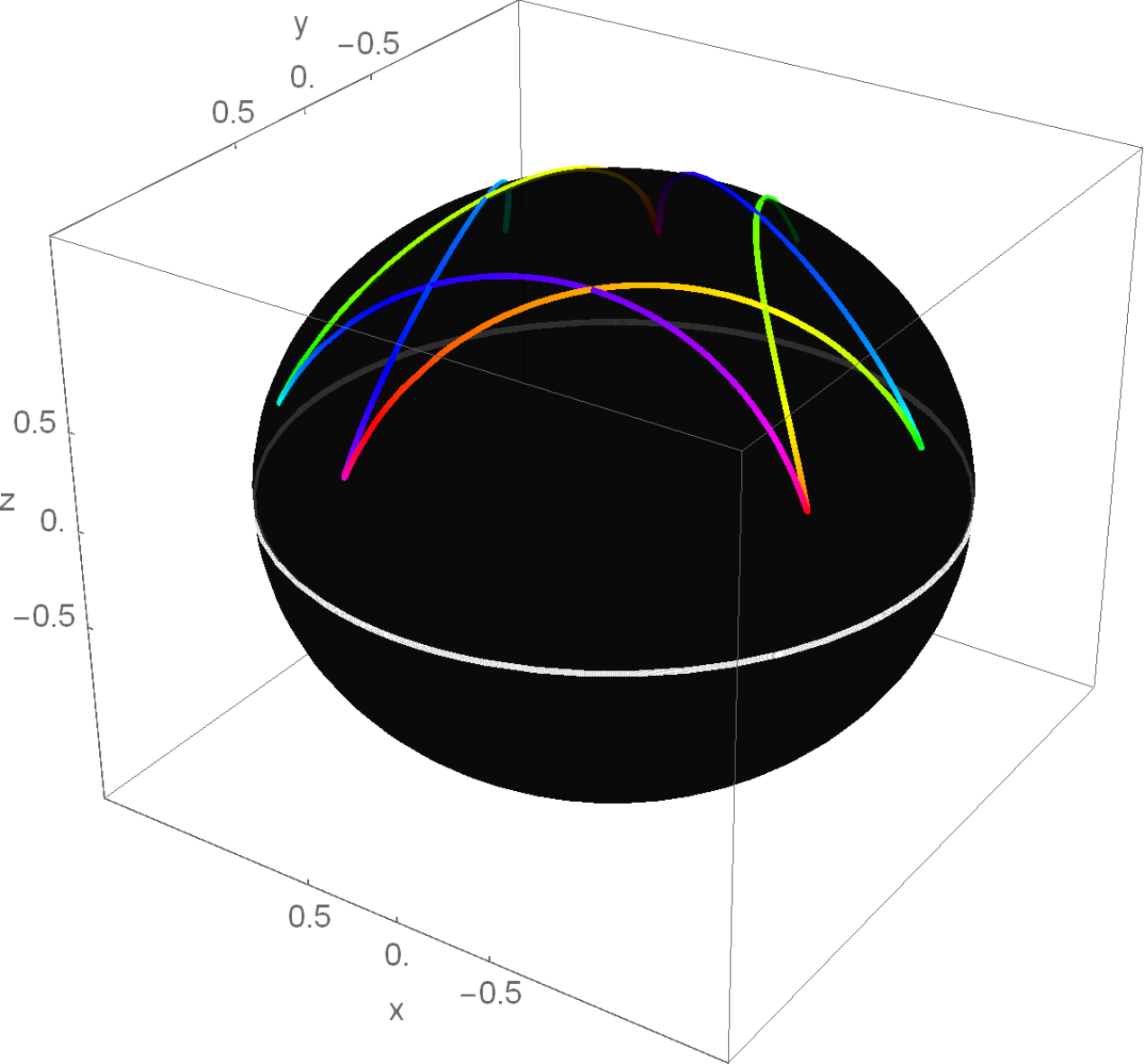}
	\caption{An orbit with $\Phi = \Psi = 0.8$ around an extremal black hole with rotation parameters $a=0.9$ and $b=0.1$.}
	\label{fig:cusp}
\end{figure}

\begin{figure}[h]
 	\includegraphics[width = 8.4cm]{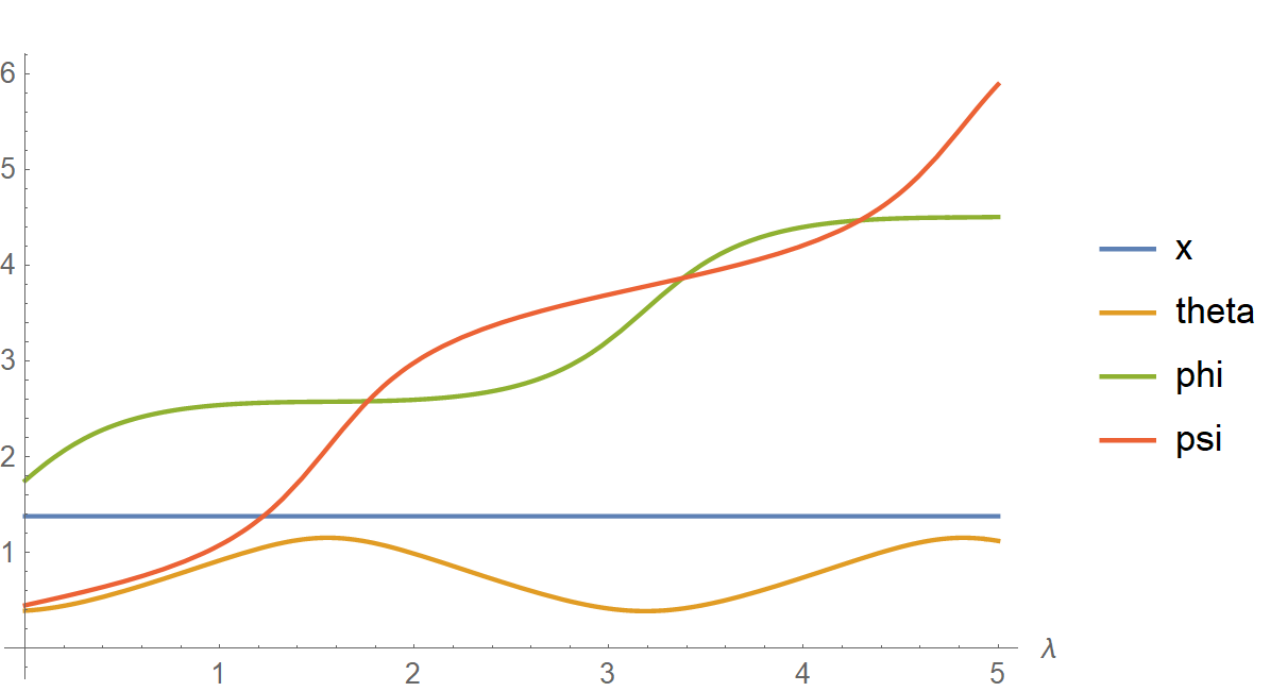}
 	\caption{The numerical solution for the coordinates along the geodesic in Figure \ref{fig:cusp}, as a function of the affine parameter $\lambda$.}
 	\label{fig:lambda}
 \end{figure}

The example in  Figure \ref{fig:cusp} shows the appearance of cusps in the geodesic orbit. Although this seems to contradict the differentiability of the geodesic, the cusps are actually just an artifact of the projection down to 3 dimensions. The colour of the geodesic changes sharply at each of the cusps, indicating that the motion of the geodesic is almost entirely in the $\psi$ direction at that point. This is clear if we plot the angular coordinates as a function of the affine parameter $\lambda$, as done in Figure \ref{fig:lambda}. One can see that when $\theta$ is at a local maximum, $\phi$ is almost constant, but $\psi$ is changing rapidly.
 
\begin{figure}[h]
	\includegraphics[width = 8.4cm]{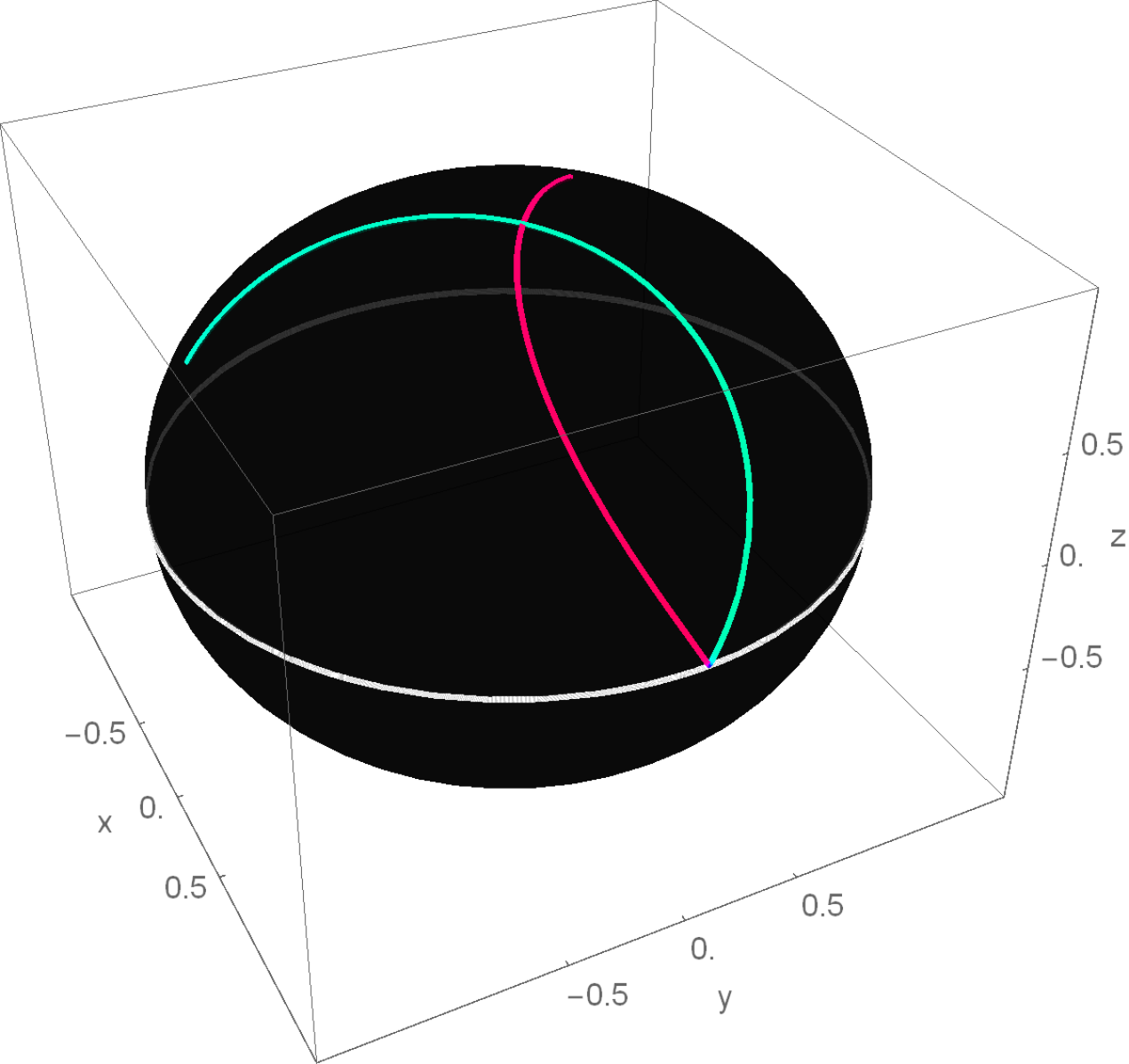}
	\caption{An orbit with $\Phi = 0.1$ and $\Psi = 0.01$ around a black hole with rotation parameters $a = 0.9$ and $b = 0.09$.}
	\label{fig:confused}
\end{figure}

\begin{figure}[h]
	\includegraphics[width = 8.4cm]{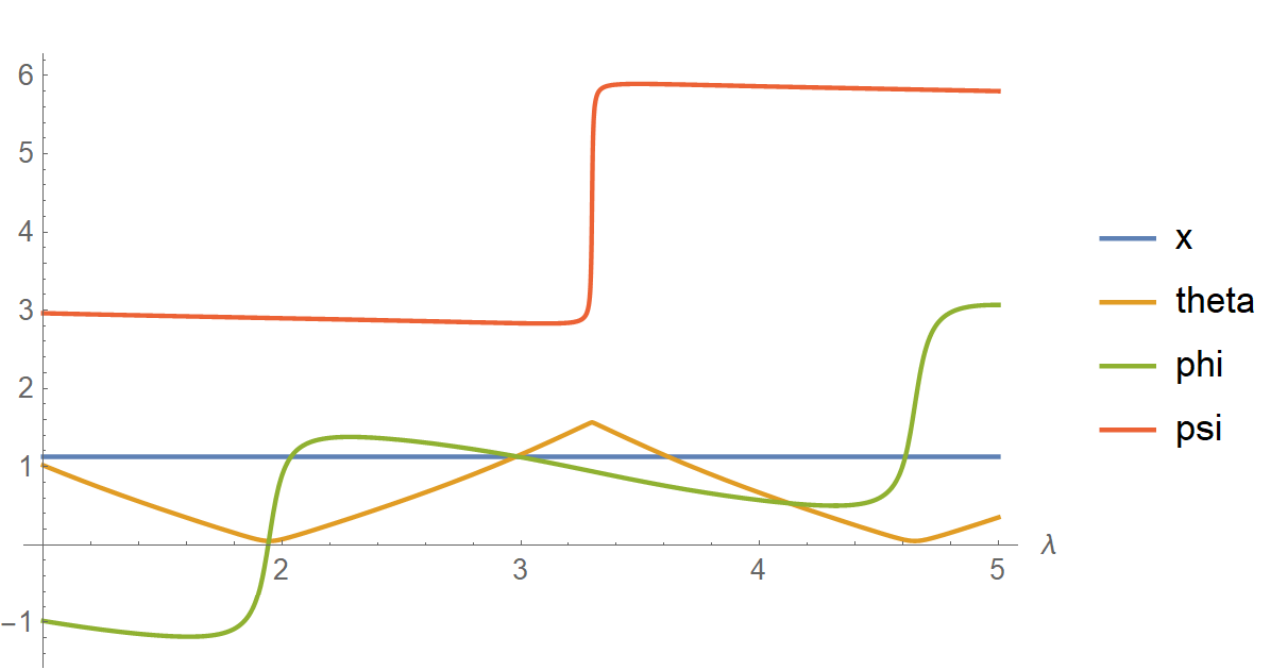}
	\caption{The numerical solution for the coordinates along the geodesic in Figure \ref{fig:confused}, as a function of the affine parameter $\lambda$.}
	\label{fig:lambda2}
\end{figure}

The penultimate example, shown in Figure \ref{fig:confused}, is an orbit that does not have a fixed azimuthal direction - that is, the coordinate $\phi$ is not monotonically increasing along the geodesic. In addition, the motion in the $\psi$ direction is also not monotonically increasing along the geodesic, which is perhaps clearer when looking at the plot of the individual coordinates as a function of the affine parameter $\lambda$ (Figure \ref{fig:lambda2}). 

As we can see, when the geodesic is close to the equator (i.e. $\theta$ is at a maximum), the $\phi$ coordinate is decreasing along the geodesic. When the geodesic is close to the pole, however, the $\phi$ coordinate is increasing along the geodesic. The $\psi$ coordinate exhibits the opposite behaviour, increasing near the equator and decreasing near the poles. This behaviour can be attributed to the five-dimensional version of the Lense Thirring effect. When $\theta$ is close to zero, the first term in (\ref{geodesicphi}) dominates, so $\phi$ is rapidly increasing at that point. When $\theta$ is close to $\frac{\pi}{2}$, the first term is dominated by the other two terms (since $\Phi$ is small), and so $\phi$ is decreasing at that point. The small, positive angular momentum keeps the geodesic moving in the positive $\phi$ direction until it gets near the equator (that is, the $x_1 - y_1$ plane), where the dragging of intertial frames in the $\phi$ direction is strongest. Similarly, the small, positive angular momentum in the $\psi$ direction keeps the geodesic moving in the positive $\psi$ direction until it gets near the pole (i.e. the $x_2 - y_2$ plane), where the dragging of inertial frames in the $\psi$ direction is strongest.   

The final example we will consider is a relatively simple one - it is a circular orbit of constant $\theta$. This orbit is a slight generalisation of the types of orbit discussed in Section \ref{subsec:equatorial}. As with all other examples mentioned in this paper, it is an unstable orbit.

 \begin{figure}[h]
	\includegraphics[width = 8.4cm]{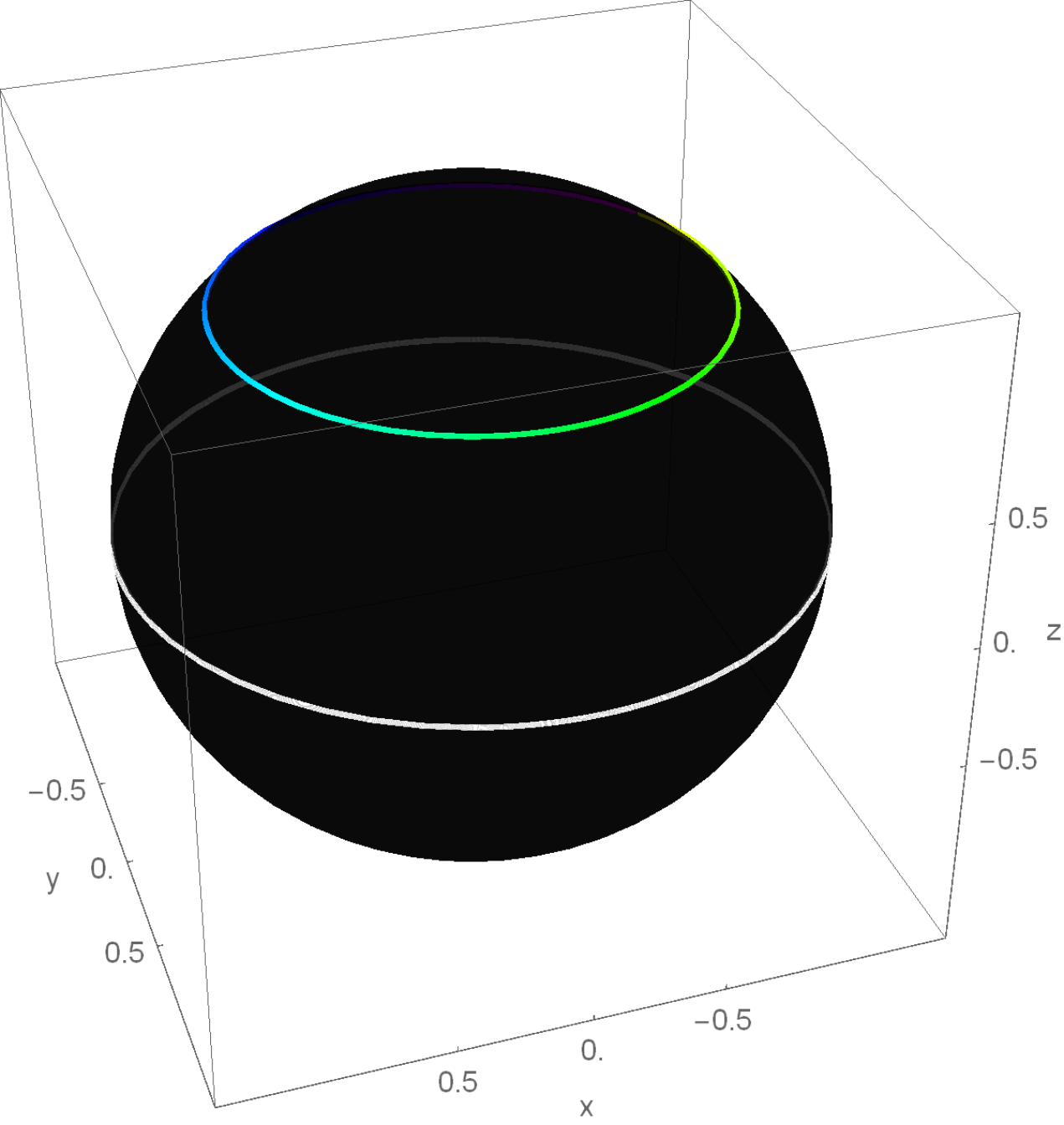}
	\caption{A black hole with rotation parameters $a = 0.5$ and $b = 0.25$. The parameters $\Phi$ and $\Psi$  were chosen so that the orbit lies at $\theta = \frac{\pi}{4}$. They are approximately $\Phi = -0.9292$ and $\Psi = 0.9036$.}
	\label{fig:boring}
\end{figure}

\clearpage

\section{Summary and Conclusions}
\label{sec:summary}
In this paper we have studied spherical photon orbits outside the event horizon of a 5D Myers-Perry black hole. We have found necessary conditions for the existence of these orbits in terms of the parameters $a,\,b,$ and $\mu$ of the black hole, as well as the conserved quantities $\Phi$ and $\Psi$, and we have also derived analytically the radii of circular null orbits in the equatorial and polar planes. We then numerically integrated the geodesic equations for an appropriate selection of variables, and plotted the results. Our method of visualising the photon orbits used a transformation to three-dimensional pseudo-cartesian coordinates, with a suppressed angular coordinate being encoded with the use of colour. Some qualitative features of orbits we found include the appearance of cusps, orbits with non-fixed $\phi$ or $\psi$ direction attributable to a five-dimensional Lense-Thirring effect, a zero momenta orbit with constant $\phi$ and a zero momenta orbit with constant $\psi$, and a non-equatorial orbit with constant $\theta$. In all cases these geodesics exhibit interesting orbital dynamics. 
 \section*{Acknowledgements}
I would like to thank the Institut Henri Poincar\'{e} for their hospitality during the early parts of this project, and I would like to thank Kyle Wright for his comments on a draft version of this paper. Finally, I would like to thank Claudio Paganini for many insightful discussions, as well as his comments on a draft version of this paper. This research was supported by the Australian government through an Australian Postgraduate Award


\end{document}